# Fabrication of photonic nanostructures from hexagonal boron nitride

*Johannes E. Fröch,[1,†], Yongsop Hwang[2,†], Sejeong Kim[1,*], Igor Aharonovich[1], Milos Toth[1]*

1. School of Mathematical and Physical Sciences, University of Technology Sydney, Ultimo, New South Wales, 2007, Australia
2. School of Engineering, RMIT University, Melbourne, VIC 3001, Australia

*† These authors contributed equally to this work*
*\* sejeong.kim-1@uts.edu.au*

**ABSTRACT**

Growing interest in devices based on layered van der Waals (vdW) materials is motivating the development of new nanofabrication methods. Hexagonal boron nitride (hBN) is one of the most promising materials for studies of quantum photonics and polaritonics. Here, we report in detail on a promising nanofabrication processes used to fabricate several hBN photonic devices using a hybrid electron beam induced etching (EBIE) and reactive ion etching (RIE) technique. We highlight the shortcomings and benefits of RIE and EBIE and demonstrate the utility of the hybrid approach for the fabrication of suspended and supported device structures with nanoscale features and highly vertical sidewalls. Functionality of the fabricated devices is proven by measurements of high quality cavity optical modes ($Q$~1500). Our nanofabrication approach constitutes an advance towards an integrated, monolithic quantum photonics platform based on hBN and other layered vdW materials.

## INTRODUCTION

Since the advent of two-dimensional (2D) materials, their peculiar physical properties have been studied intensively, due to their potential applications as atomically thin integrated devices [1-5]. Novel fabrication protocols were developed to reliably process and functionalize these materials, and enable control of their physical, optical, and electronic properties at the nanoscale. In the field of photonics, hexagonal boron nitride (hBN) emerged as a material of interest because it is the first known naturally hyperbolic material with a highly anisotropic dielectric permittivity which allows directional, sub-diffraction guided modes [6-8]. Additionally, hBN has a wide bandgap (~6 eV)[9] which provides a transparent optical platform in the ultraviolet-to-near-infrared range, and hosts a range of atomic point defects that act as ultrabright, photostable, room temperature single photon emitters (SPEs)[10-14].

SPEs have numerous potential applications in integrated, on-chip quantum photonic circuits and quantum information processing [15-17]. Integration of SPEs with dielectric cavities (such as microdisk or microring cavities, photonic crystal cavities) is an important step towards the realization of integrated quantum photonic circuits. Initial results on integration of SPEs from other layered materials such as $WSe_2$ have been demonstrated by integrating them into 1D silicon nitride cavities or open cavities [18-19]. On the other hand, hBN can be used in an entirely monolithic approach where the devices are fabricated from the parent material that hosts SPEs[20-22]. The monolithic approach is advantageous in that emitters can be located within the high energy field of the cavities, and can therefore be positioned in the maxima of optical modes, thus enabling the realisation of optimal coupling efficiencies as previously done for diamond, GaAs, and SiC[23-26]. This in combination with recent reports on the tunability of the SPE emission wavelength will allow to directly tune emitters to the resonant modes of devices[27], in order to reversibly investigate effects such as Purcell enhancement from hBN SPEs.

In this work we report detailed nanofabrication protocols used to realize photonic resonators from hBN. In particular, we demonstrate microring cavities and 1D photonic crystals that exhibit high quality factors ($Q$) in excess of 1,000. We also discuss in detail the shortcomings and benefits of a hybrid RIE-EBIE process that is suitable for the fabrication of large-area, suspended and supported device structures that are made from hBN, and contain nanoscale features and highly vertical sidewalls. The nanofabrication approach presented here is suitable for other layered materials, and therefore broadly relevant to the field of photonics, as well as polaritonic -, nanoelectromechanical -, optoelectronic -, and optomechanic systems[28], all of which require engineering of structures with nanoscale precision.

**RESULTS AND DISCUSSION**

The fabrication process is outlined schematically in Fig.1. First, hBN flakes are transferred onto a silicon or a silicon dioxide substrate via mechanical exfoliation using sticky tape (Fig. 1a). hBN has the advantage that it can be exfoliated from a larger parent crystal, to flakes that are chemically and mechanically stable/robust, with thicknesses as small as a single monolayer. Therefore, it does not require cumbersome preparation steps that are often necessary for classical bulk counterparts such as diamond and SiC. The substrate was pre-patterned with trenches that are ~ 5 μm deep to allow for a sufficient air gap below the suspended hBN structures. After hBN transfer (Fig. 1b), residues and contaminants from the sticky tape were removed by calcination in air on a hot plate at 500 $^0$ C and subsequent annealing in Argon at 850 $^0$ C, which also increases adhesion of hBN flakes to substrates. An optical image of a suspended flake before the subsequent processing steps is shown in Fig. 1c. Next, a 15 nm tungsten film was deposited to serve (later on in the process) as a mask for electron beam induced etching (EBIE) [29-30]. The substrate was spin coated with a single layer of PMMA A4 (5000 rpm, 50 s) (Fig. 1d), and nanophotonic structures were patterned into the resist using conventional electron beam lithography (EBL) at an electron beam energy of 20 keV and a beam current of 20 pA. The patterns were then transferred from PMMA into the tungsten layer and hBN by reactive ion etching (RIE) using 10 mTorr of $SF_6$, a flow rate of 60 SCCM, a power of 100 W, a bias of 300 V and a processing time of ~ 30 s (Fig. 1e). After RIE, the PMMA which served as the RIE etch mask, was removed using acetone and the substrate was additionally treated by $O_2$ RIE ashing (50 W, 150 V Bias, 60 SCCM, 10 mTorr) in order to remove PMMA residuals. An optical image of the flake after RIE is shown in Fig. 1f, where it can be seen that the patterns were transferred into the W mask and the underlying hBN, although the flake was not penetrated entirely in this etching step. After RIE, we used EBIE (1 keV, 10 nA) in an environmental SEM (Fig. 1g) with water vapor as the reactive agent at a chamber pressure of 100 mTorr to further etch the flakes[30]. The patterned tungsten served as an etch mask in this step. Monte Carlo simulations of electron-solid interactions (performed using the package CASINO[31]) were used to determine that a 15 nm tungsten layer is a sufficient mask for hBN processed with a 1 keV electron beam, avoiding penetration of the beam through the mask and undercutting of the mask. Later, we demonstrate that a single RIE processing step is a possible alternative to the hybrid RIE-EBIE method, but the EBIE step is beneficial as it improves sidewall verticality and process resolution. After all processing steps, the remaining tungsten is removed by a $H_2O_2$ (30 %) solution (Fig. 1h). An optical image of final suspended structures (Fig. 1i) shows that they are etched throughout the entire thickness of the hBN flake. We note here that for fabrication of supported devices, as presented later, a thicker PMMA layer (2500 rpm, 50 s) was used, allowing for longer RIE processing times (~ 45 s), which would result in etching of the underlying substrate ($SiO_2$) in less confined areas. This yielded small $SiO_2$ pedestals that improve the refractive index contrast of the devices. However, we still employed an additional EBIE step, in order to achieve vertical sidewalls and to etch smaller confined featured.

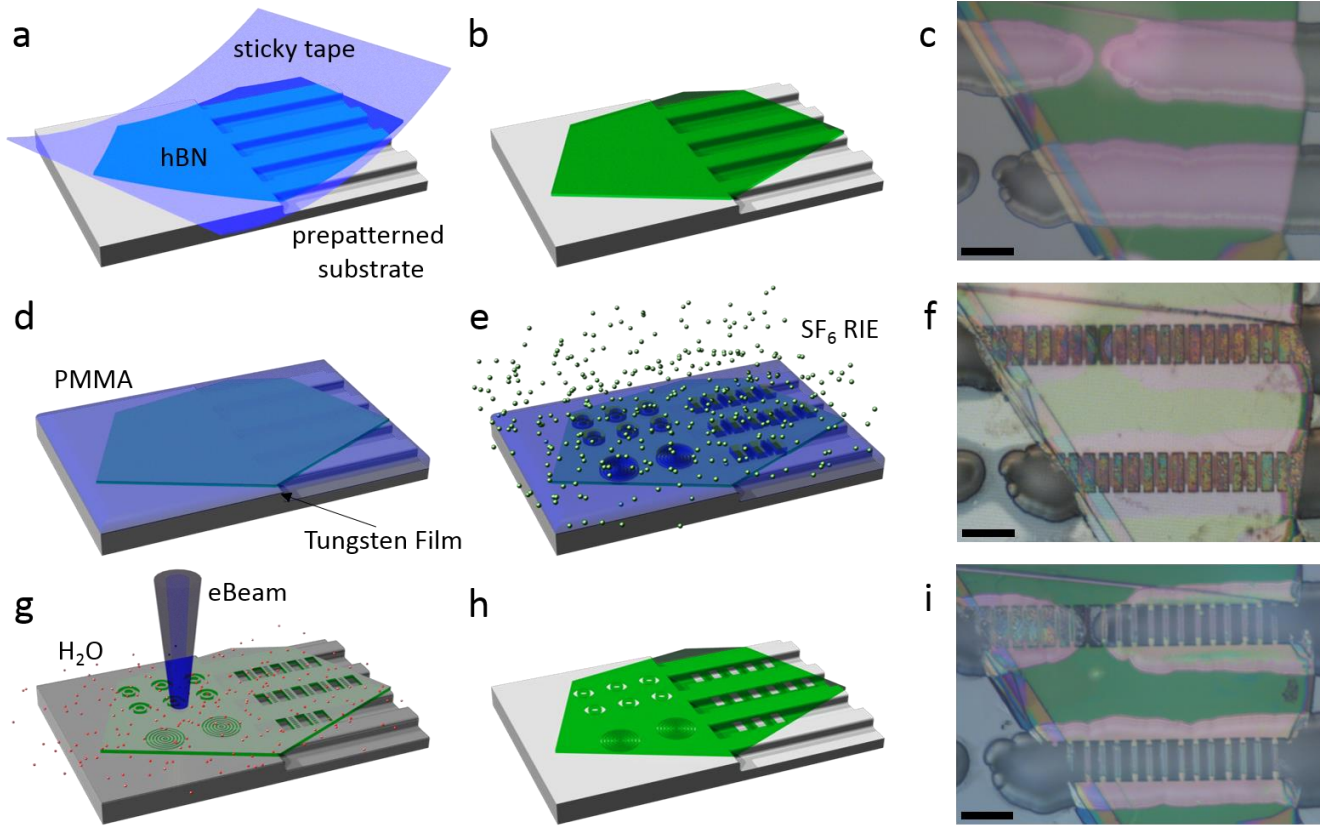

**Figure 1.** Fabrication of photonic structures from hBN. (a) Mechanical exfoliation of a hBN flake from sticky tape onto a pre-patterned substrate. (b) An as-exfoliated hBN flake after a cleaning process. (c) Optical image showing such a flake before further fabrication steps. (d) Deposition of a tungsten layer and spin coating of a PMMA layer, which serve as masks in the EBIE and RIE steps, respectively. (e) E-beam lithography followed by $SF_6$ RIE. (f) Optical image of the hBN flake containing 1D photonic crystals after RIE and PMMA removal, showing that patterns are transferred into tungsten and hBN. (g) Masked EBIE used to complete the hBN etch. (h) Final hBN photonic structures after tungsten removal. (i) Optical microscope image of the hBN flake obtained after all processing steps. Scale bars in optical images correspond to 10 μm.

The developed nanofabrication process is directly applicable to arbitrary device geometries, as shown in Fig. 2. It is possible to fabricate suspended photonic devices, such as photonic crystal cavities (a) and (b), or devices supported by substrates, such as whispering gallery mode (WGM) microring resonators (c), circular grating structures (d), photonic waveguides with vertical in- and outcoupling gratings (e). Since the developed protocol does not rely on mask lift-off, patterning directly at the EBL resolution is possible without any limitation, enabling feature sizes down to few tens of nanometers. For example, in the case of a photonic crystal cavity (a) airholes with features of 100 nm are fabricated over a length of several micrometer. Sidewalls appear vertical and well resolved, directly demonstrating the high potential of this nanofabrication approach. Another distinct advantage of this technique is that no sonication or rinsing in solution is necessary for mask lift off, which can relocate smaller free lying features, such as the circular grating and the outcoupling grating of a hBN waveguide shown in (d) and (e). The devices in Fig 2. constitute the basic components of a quantum photonic chip, and demonstrate that the fabrication technique will make it possible to explore the field of integrated hBN nanophotonics by fabricating structures around existing and pre-characterized quantum emitters, since electron beam induced processing has been shown to be non-invasive to SPEs embedded in hBN [11]. In fact, some emitters in hBN have been observed to be highly resilient to a range of processing treatments, including irradiation by focused $He^+$ beams[32]. A low energy electron beam, such as utilized here, will not alter existing defects within hBN through knock-on (i.e. momentum transfer), as the energy threshold for electron knock-on damage is 88 keV, and 130 keV for B and N atoms, respectively[33].

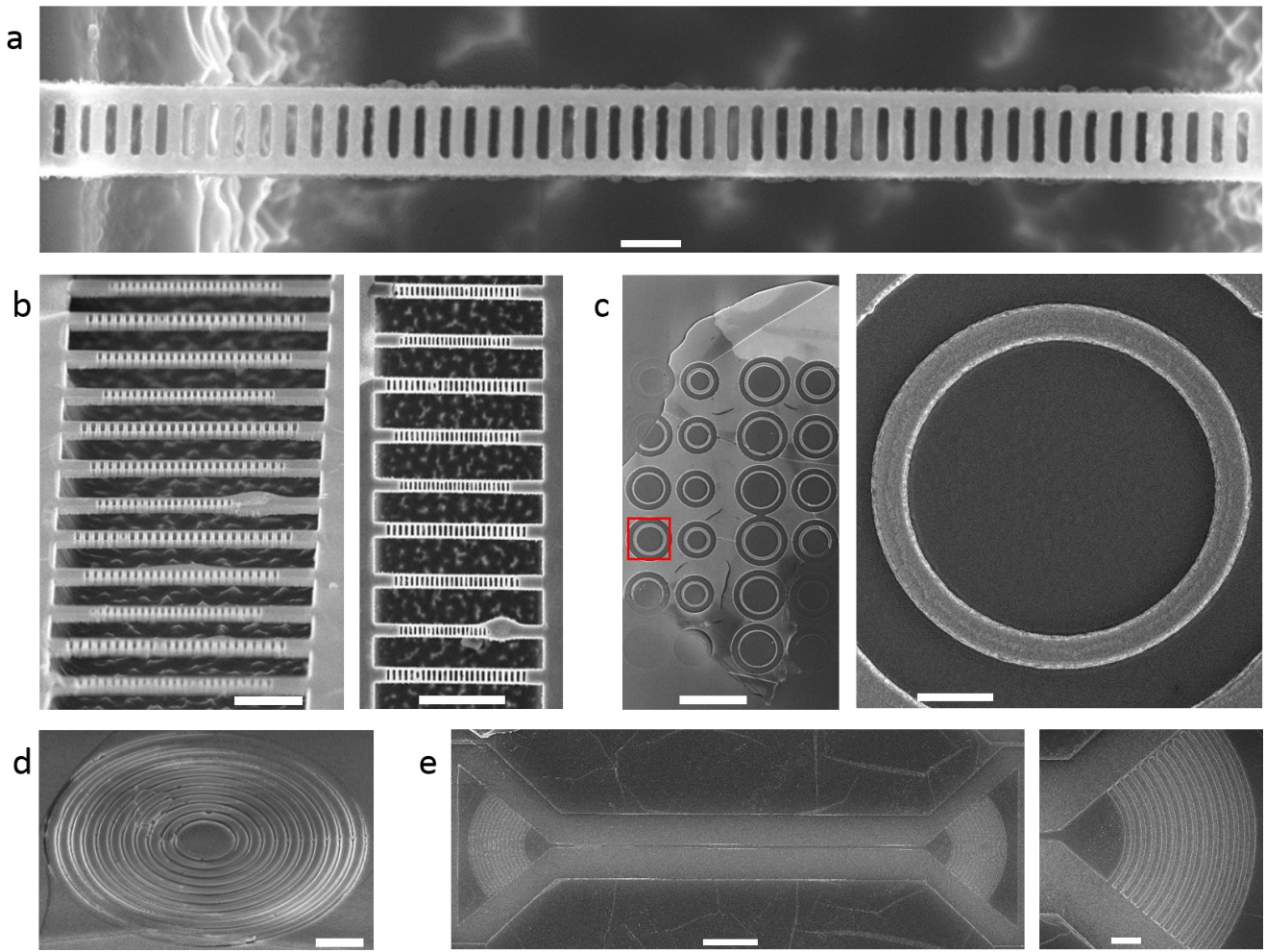

**Figure 2.** Various photonic structures fabricated from hBN. (a) Top view of a photonic crystal cavity with airhole sizes of 100 nm x 450 nm over a length of 10 μm. The scale bar corresponds to 500 nm. (b) Tilted and top view images of free-standing one-dimensional photonic crystal cavity arrays. Scale bars correspond to 2 μm and 4 μm, respectively. (c) A set of hBN microrings with various sizes and a magnified view of the highlighted area. Scale bars correspond to 10 μm and 1 μm, respectively. (d) A circular grating structure in a tilted view, the scale bar corresponds to 2 μm. (e) A waveguide with two grating couplers and its magnified image. Scale bars correspond to 5 μm and 1 μm, respectively. Structures in (c), (d) and (e) are on a $SiO_2$ substrates.

The combination of RIE and EBIE, however, suffers from several potential pitfalls that can give rise to fabrication artefacts and compromise device functionality. In Fig. 3 several of these artefacts are shown.

In the case of RIE, it is well known that contaminants in the chamber, an unsuitable mask, or a high physical sputter rate produced by unfavorable RIE conditions can lead to micromasking which gives rise to spots with a size on the order of few tens of nanometers and grassing. Similarly, we observe micromasking effects during EBIE. To illustrate, the structures in Fig 3. (a) contain spots on the surface of the masked hBN, which are visible after PMMA removal. The spots likely arise from crosslinked PMMA residuals left behind due to incomplete PMMA removal, or regions that were overexposed during EBL, and can act as micromasks and persist in nominally-unmasked regions of hBN throughout the EBIE processing step (Fig. 3 (b)) since they are relatively inert to $H_2O$ EBIE. The process of EBIE is known to be highly material-selective, and a similar effect has been reported previously. Specifically, $H_2O$ EBIE of amorphous carbonaceous films was shown to lead to the formation of nano-scale regions that are highly resistant to EBIE, and attributed to electron beam restructuring of the films via processes such as H-C bond cracking, high order carbon-carbon bond formation, and H desorption[34]. The spots seen in Fig. 3(b) inhibit EBIE of the underlying hBN, leading to the formation of hBN needles seen in the tilted-view image shown in Fig. 3 (c). To circumvent this issue, the dosage used for EBL must be optimised. Moreover, it is necessary to use low electron beam currents to avoid overexposing of the resist whilst imaging when the flake is positioned for EBL. After $SF_6$ RIE, it is important to ensure that the substrate is cleaned to remove residual PMMA using both acetone and an additional $O_2$ RIE ashing step.

Similarly to the previous effect, we observed that the presence of hydrocarbon contaminants in the SEM chamber can lead to the formation of features that are not etched by $H_2O$ EBIE. Most likely, some contaminants react with hBN during electron exposure, forming a compound that is inert to electron beam induced etching in a water vapor atmosphere. This is illustrated in Fig. 3 (d) and (e) for areas processed with total electron doses of 33.6 C $cm^{-2}$ and 67.2 C $cm^{-2}$, respectively - such features are typically not removed by EBIE. More prolonged EBIE can cause the unetched features to agglomerate, wrap around cavity structures, as is illustrated by Fig. 3 (f), and persist even after extended processing times (~ 100 C $cm^{-2}$). The extended processing times compromise cavities due to irradiation by secondary electrons and backscattered electrons emitted from the underlying substrate. In order to avoid agglomeration and structure wrapping, the $SF_6$ RIE step has to be maximized until large areas of the hBN film are etched through. In order to suppress etching from secondary and backscattered electrons, the EBIE step has to be stopped immediately, after all features are etched. Despite the above complications, the EBIE processing step has several advantages. Images comparing the same photonic crystal cavity after RIE and after an additional EBIE step are shown in Fig. 3 (g) and (h), respectively. Although larger areas of the hBN film are fully etched by RIE, the side walls are not fully vertical and the smallest features are not fully developed in the sense that the film is not completely etched through, as is seen clearly in the enlarged view of the beam end shown in Fig. 3 (i). To date, it was reported that for an $SF_6$ RIE-based etch process, a maximum sidewall angle of $80^0$ was achieved after optimization, and only for relatively large feature sizes[35]. In contrast, the additional EBIE step utilized here yields straight sidewalls (i.e. sidewall angles of $\sim 90^0$) and smaller features are etched thoroughly, as is illustrated by the image shown in Fig. 3 (j). The shortcomings of RIE are attributed to the fact that the plasma needs to penetrate the etched regions of hBN, and material removed by the physical sputtering component of the RIE process can redeposit on surrounding sidewalls. In contrast, the electron beam can easily penetrate nanoscale holes, and the etch process is chemical, thereby mitigating redeposition artefacts under clean conditions. We also note that $H_2O$ EBIE of materials such as single crystal diamond and hBN has previously been shown to be minimally invasive in the sense that Raman spectra of the processed materials do not show evidence of structural damage that is typically produced by ion bombardment/etching [30,36].

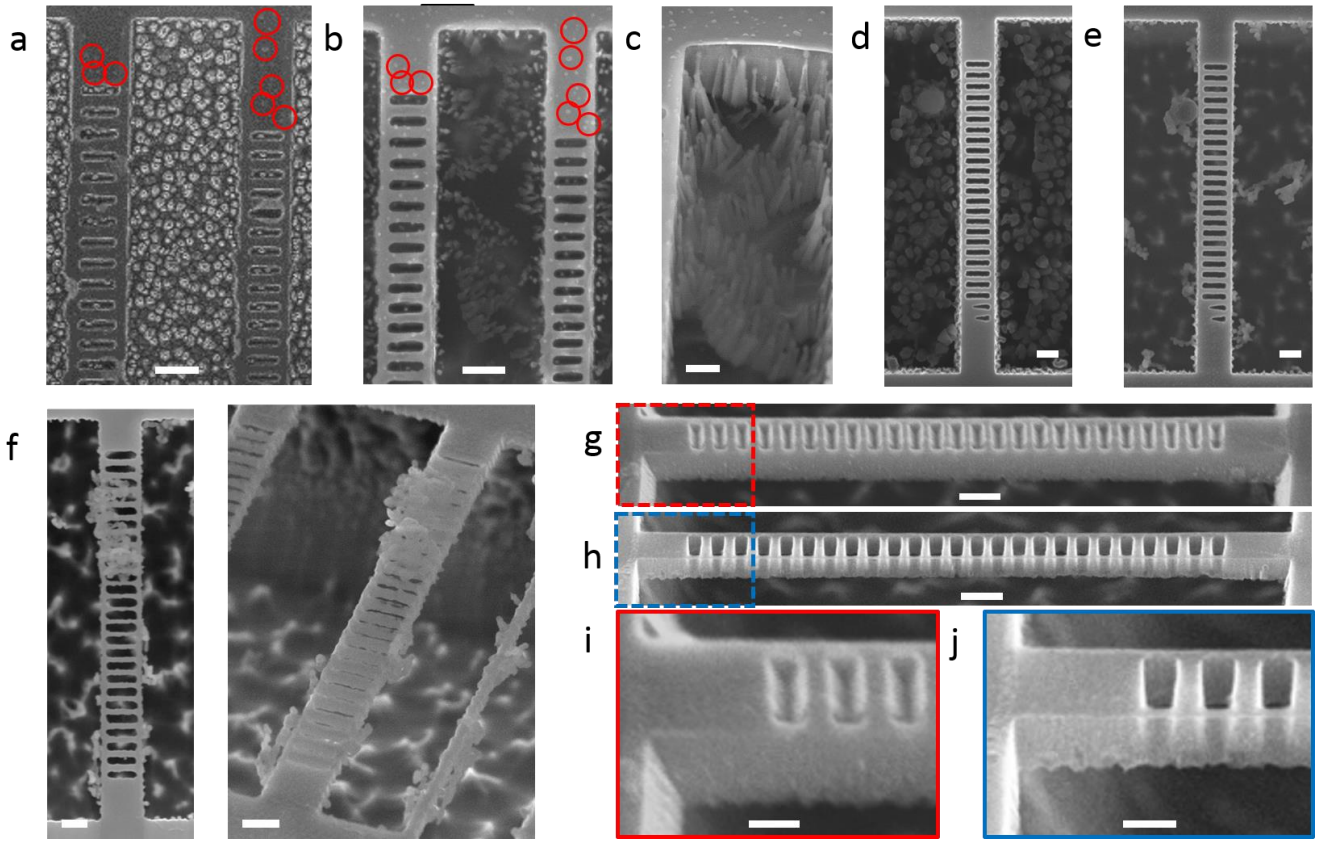

**Figure 3.** Fabrication artefacts of hBN. Photonic crystal cavities imaged after RIE (a), showing small spots from micromasking. Some spots are highlighted with red circles. The same spots persist after EBIE (b), and give rise to the formation of grass in nominally unmasked regions of hBN, seen clearly in the tilted view in (c). Scale bars correspond to 500 nm in (a) and (b), and 250 nm in (c), respectively. Two images of the same nanobeam obtained after subsequent EBIE steps, (d) and (e), show that some areas of unmasked hBN are not etched, producing features that agglomerate and wrap around the cavities. Scale bars correspond to 500 nm. An overetched structure (f) shown in top and tilted views, in which hBN residuals are not removed even after prolonged EBIE. The integrity of the photonic crystal cavity is compromised by agglomeration of residuals. Scale bars correspond to 500 nm. A comparison of the side view after RIE only (g) and after the subsequent EBIE processing step (h). Scale bars correspond to 500 nm. (i) and (j) are enlarged views of the outlined sections in (g) and (h). Scale bars correspond to 250 nm.

We now turn to characterization of the devices using photoluminescence (PL) measurements, carried out using a home built confocal microscope setup at room temperature with a 532 nm excitation source and 300 µW excitation power. For a suspended nanobeam cavity, a resonance at ~ 701 nm, with a $Q = \lambda/d\lambda \sim$ 1,500 is observed, as shown in Figure 4a, b (figure 4b is a close-up of the cavity mode seen in 4a). The inset of Fig 4a shows an SEM image of the cavity, with a design discussed elsewhere[22]. For a supported device, (i.e. a resonator that rests on a substrate) we demonstrate WGMs of a microring resonator atop an $SiO_2$ substrate (Fig. 4c). Figure 4c shows the WGMs (periodic peaks in the spectrum) from the hBN microring. Figure 4d shows a higher resolution spectrum of one of the WGMs, with a measured $Q \sim 500$ at a wavelength of 644 nm. Although the $Q$ of this device is rather moderate, it can be improved in the future by using a substrate with a lower refractive index, such as $CaF_2$, $MgF_2$ or an aerogel.

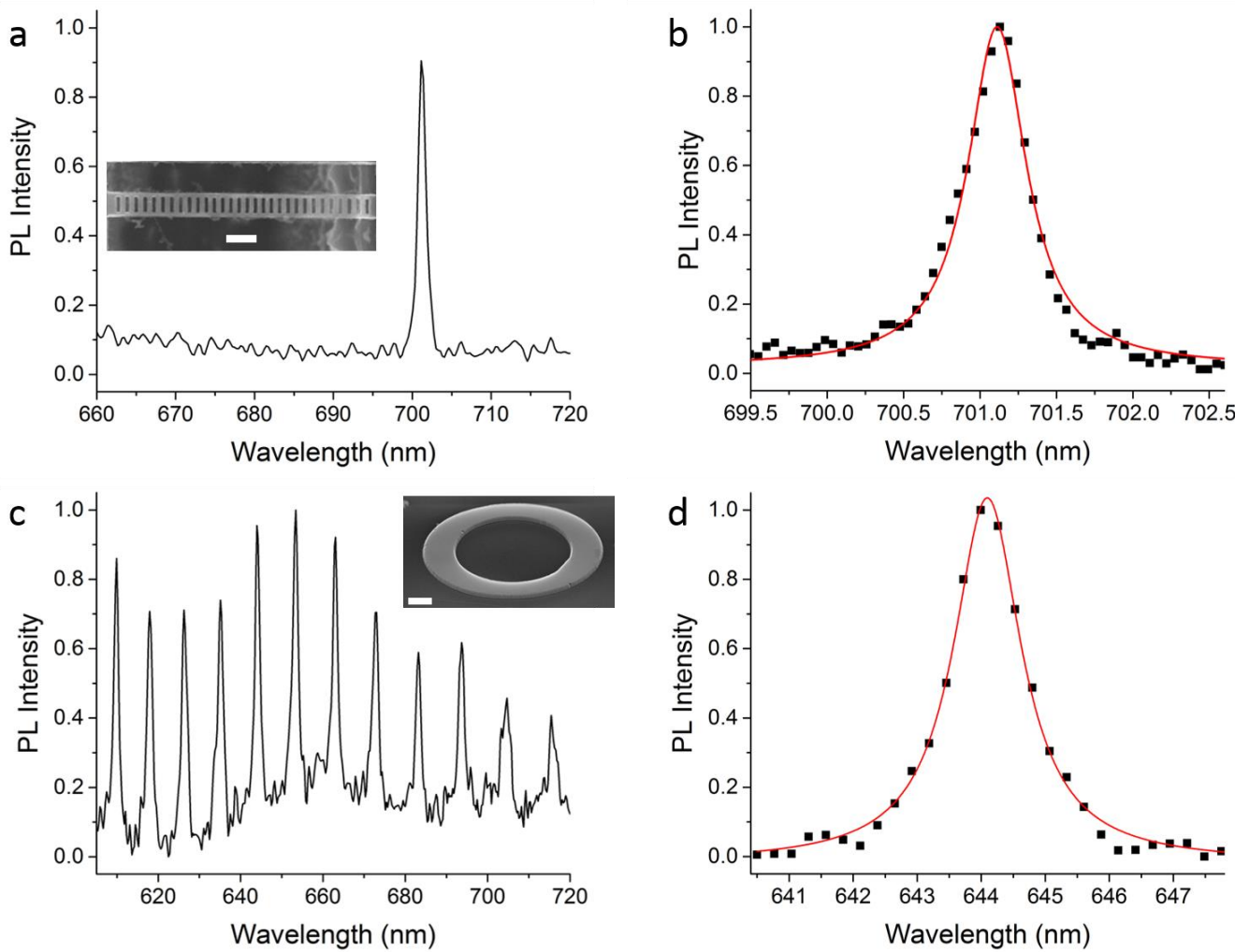

**Figure 4.** Quality factors of the fabricated hBN devices. (a) A 1D PCC with a resonant mode at ~ 701 nm. The inset shows the measured device, with a scale bar of 1 μm. (b) The cavity $Q$ of ~ 1500 is determined from a Lorentzian fit at the resonant wavelength. (c) Whispering gallery modes of a hBN microring. The inset shows a tilted view of the device. The scale bars correspond to 1 μm. (d) The $Q$ of the microring resonator was determined to be 500 by a Lorentzian fit of a resonance at 644 nm.

**CONCLUSION**

We detailed step-by-step fabrication of suspended and supported photonic devices from hBN. For the developed nanofabrication approach, we utilize a combination of RIE and EBIE. We emphasize the versatility of this process by demonstrating several fundamental photonic devices including microdisk cavities and photonic crystals, which will enable coupling of hBN SPEs to cavity modes to explore the field of integrated hBN quantum photonics. For a better understanding of the fabrication process, we demonstrate the cause of several fabrication errors, which we observed during the development of this technique and could mainly relate to contamination arising either from the RIE or EBIE step. The best structures achieved so far were realized by maximizing the RIE step, and by using EBIE as a final processing step that improves resolution and sidewall verticality. The presented fabrication approach is not limited to hBN and can be extended to other layered materials. It is therefore of general relevance to the fields of photonics, optoelectronics, optomechanics, and nanoelectromechanical systems.

**ACKNOWLEDGEMENTS**

Financial support from the Australian Research council (via DP140102721, DP180100077, LP170100150), the Asian Office of Aerospace Research and Development grant FA2386-17-1-4064, the Office of Naval Research Global under grant number N62909-18-1-2025 are gratefully acknowledged.